\documentclass[11pt]{article}
\usepackage{graphicx}

\newcommand{\BABARPubYear}    {02}

\newcommand{\BABARConfNumber} {19}
\newcommand{\SLACPubNumber} {9322}

\input pubboard/babarsym
\newcommand{\btoddk}{\ensuremath{B\ra \overline{D}^{(*)} D^{(*)} K}}
\newcommand{\de}{\ensuremath{\Delta E}}

\newcommand{\bdzdzk}{$B^0\rightarrow \overline D^{0}D^{0}K^+$}

\newcommand{\btodsdsk}{\ensuremath{B^+ \to D^{*-} D^{*+} K^+}}

\long\def\inst#1{\par\nobreak\kern 4pt\nobreak
    {\it #1}\par\vskip 10pt plus 3pt minus 3pt}

\begin{document}
{\pagestyle{empty}
\begin{flushright}
\babar-CONF-\BABARPubYear/\BABARConfNumber \\
SLAC-PUB-\SLACPubNumber \\
July 2002 \\
\end{flushright}

\par\vskip 5cm

\begin{center}
\Large \bf Measurement of the Branching Fractions for the Exclusive Decays of 
 \boldmath ${B^0}$ and \boldmath ${B^+}$ to 
\boldmath ${\overline D^{(*)}D^{(*)}K}$
\end{center}
\bigskip

\begin{center}
\large The \babar\ Collaboration\\
\mbox{ }\\
July 26, 2002
\end{center}
\bigskip \bigskip

\begin{center}
\large \bf Abstract
\end{center}
Using data collected with the \babar\  detector between 1999 and 2002, we 
report the observation of  823$\pm$57 $B^0$ and 969$\pm$65 $B^+$  decays 
to $\overline{D}^{(*)}D^{(*)}K$, where $\overline D^{(*)}$ and $ D^{(*)}$ 
are fully reconstructed and where $K$ is either a $K^\pm$ or a $K^0_S$ 
decaying to $\pi^+\pi^-$. All 22 possible $B$ decays to 
$\overline{D}^{(*)}D^{(*)}K$ are reconstructed exclusively and the 
corresponding branching fractions or limits are determined. 
The preliminary branching fractions of the $B^0$ and of the $B^+$ to 
$\overline D^{(*)} D^{(*)} K$ are found to be   
\begin{center}
${\cal B}(B^0\rightarrow \overline D^{(*)} D^{(*)} K) = \left (4.3 \pm 0.3 (stat) \pm 0.6 (syst) \right )\times 10^{-2}$, 
${\cal B}(B^+\rightarrow \overline D^{(*)} D^{(*)} K) = \left (3.5 \pm 0.3(stat) \pm 0.5 (syst) \right )\times 10^{-2}$.
\end{center}
 

\vfill
\begin{center}
Contributed to the 31$^{st}$ International Conference on High Energy Physics,\\ 
7/24---7/31/2002, Amsterdam, The Netherlands
\end{center}

\vspace{1.0cm}
\begin{center}
{\em Stanford Linear Accelerator Center, Stanford University, 
Stanford, CA 94309} \\ \vspace{0.1cm}\hrule\vspace{0.1cm}
Work supported in part by Department of Energy contract DE-AC03-76SF00515.
\end{center}

\newpage
} 

\begin{center}
\small

The \babar\ Collaboration,
\bigskip

B.~Aubert,
D.~Boutigny,
J.-M.~Gaillard,
A.~Hicheur,
Y.~Karyotakis,
J.~P.~Lees,
P.~Robbe,
V.~Tisserand,
A.~Zghiche
\inst{Laboratoire de Physique des Particules, F-74941 Annecy-le-Vieux, France }
A.~Palano,
A.~Pompili
\inst{Universit\`a di Bari, Dipartimento di Fisica and INFN, I-70126 Bari, Italy }
% ...

\end{center}\newpage

\setcounter{footnote}{0}
\section{Introduction}
\label{sec:Introduction}
The inconsistency between the measured $b \to c \overline c s$ rate and the rate of semileptonic $B$ decays has been a long-standing problem in $B$ physics. 
Until 1994, it was believed that the $b \to c \overline c s$ transition was 
dominated by decays $B \rightarrow D_s X$, with some smaller contributions 
from decays to charmonium states and to charmed strange baryons.  Therefore, 
the  branching fraction $b \to c \overline c s$  was computed from the inclusive 
$B \to D_s \, X $, $B \to (c \overline c)\, X $ and $B \to \Xi_c \, X $ branching 
fractions, leading  to  
${\cal B}(b \rightarrow c \overline c s)=(15.8 \pm 2.8)\%$ \cite{browder2}.
Theoretical calculations are unable to simultaneously describe this low 
branching fraction and the semileptonic branching fraction of the $B$ 
meson \cite{bigi}. 
\par As a possible explanation of this problem, it has been 
conjectured \cite{buchalla} that ${\cal B}(b \to c \overline c s)$ is in fact 
larger and that decays of the type $B \to \overline D^{(*)} D^{(*)} K\,(X)$ 
(where $D^{(*)}$ can be either a $D^0$, $D^{*0}$, $D^+$ or $D^{*+}$)\footnote {Charge-conjugate reactions are implied throughout this note.} 
could contribute significantly to the decay rate. 
This might also include possible decays to  orbitally-excited $D_s$ mesons, 
$B \to \overline D^{(*)} D_s^{**}$, followed by $D_s^{**}\to D^{(*)}\, K$. 
Experimental evidence in support of this picture has been published in the 
past few years. This evidence includes the measured branching fraction for 
wrong-sign $D$ production, averaged over charged and neutral B mesons, 
by CLEO \cite{cleoupv} [${\cal B}(B \to D\, X)=(7.9\pm 2.2)\% $], and 
the observation of a small number of fully reconstructed  decays 
\btoddk, both by CLEO \cite{cleoddk} and ALEPH \cite{alephddk}. More recently, \babar\  \cite{babarddk} and Belle \cite{belleddk} have released some preliminary 
conference results on the evidence for transitions $B^0 \to \overline D^{(*)0}D^{*+}K^-$ with much larger data sets.

\begin{figure}[H]
\begin{center}
\scalebox{.29}{\includegraphics{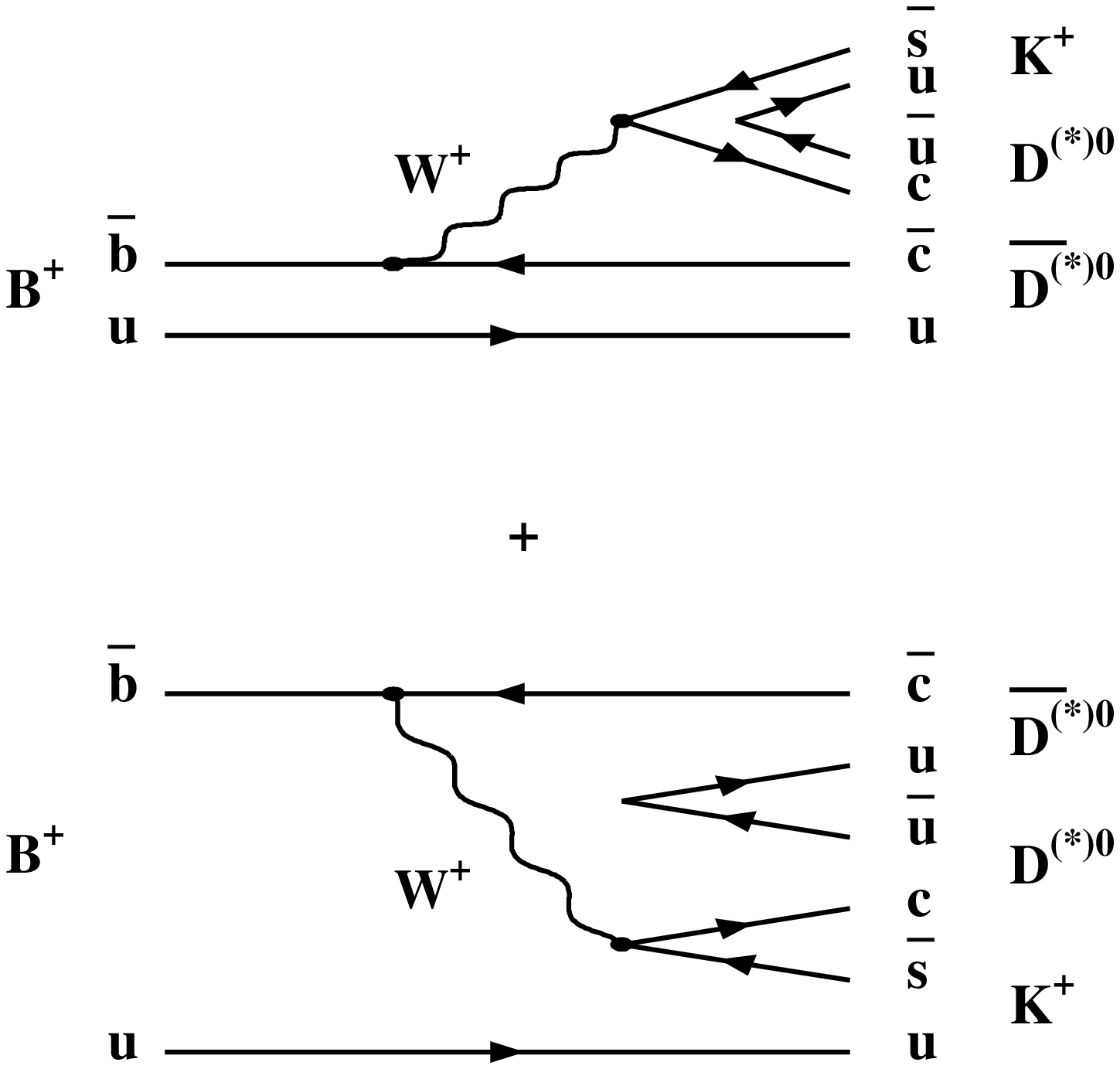}
\includegraphics{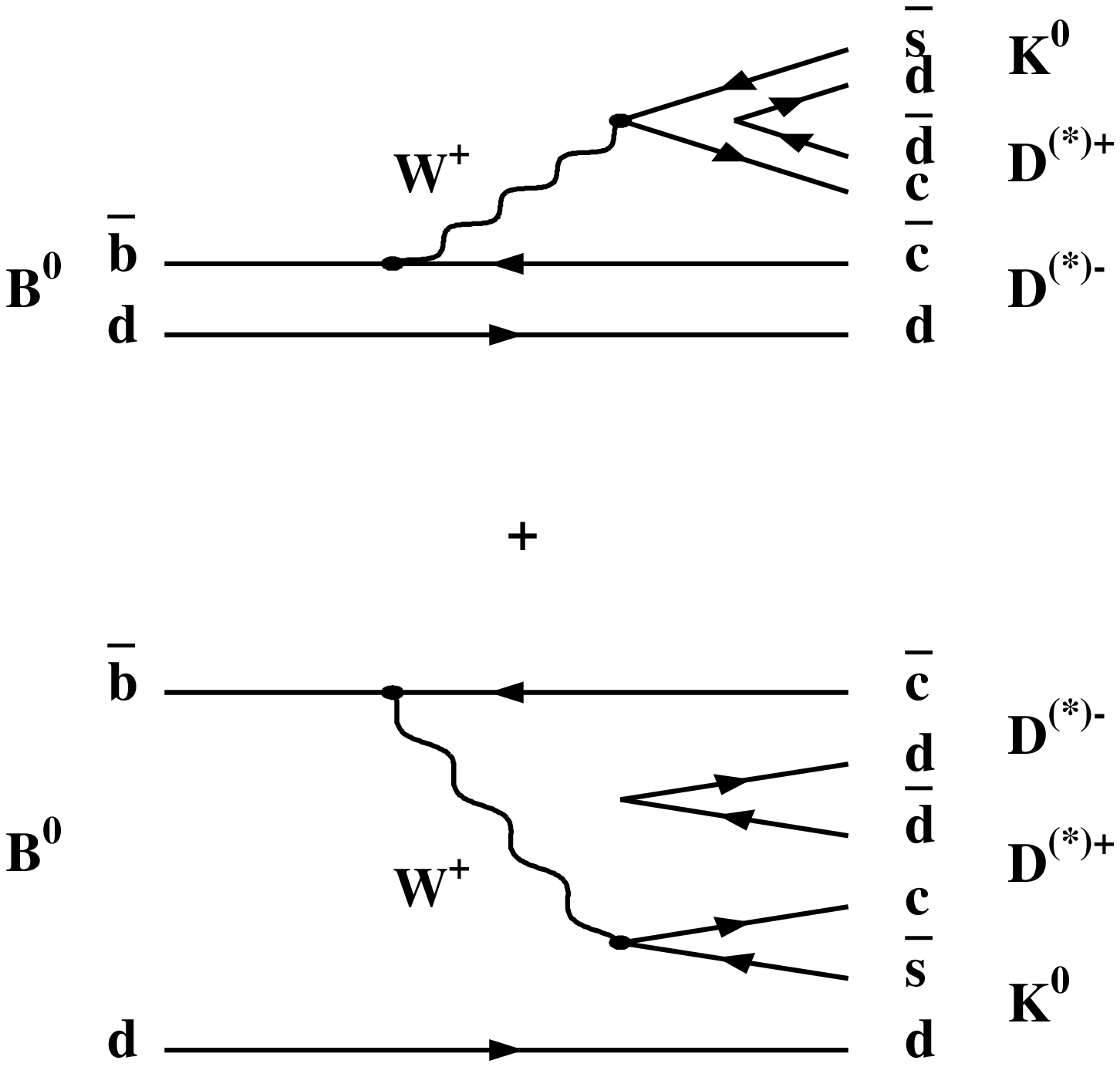}}
\caption{Top row: external W-emission amplitudes for the decays 
$B^+ \to \overline D^{(*)0} D^{(*)+} K^0$ and 
$B^0 \to D^{(*)-} D^{(*)0} K^+$. Second row:
internal W-emission amplitudes for the decays 
$B^+ \to D^{(*)-} D^{(*)+} K^+$ and 
$B^0 \to \overline D^{(*)0} D^{(*)0} K^0$. Bottom rows:
external+internal W-emission amplitudes for the decays 
$B^+ \to \overline D^{(*)0} D^{(*)0} K^+$ and 
$B^0 \to D^{(*)-} D^{(*)+} K^0$.}
\label{Fi:diagrams}
\end{center}
\end{figure}

\btoddk\ decays can occur through two different amplitudes: external 
W-emission amplitudes and  internal W-emission amplitudes (also called 
color-suppressed amplitudes). Some decays proceed purely through one of these 
amplitudes while others can proceed through both. Fig. \ref{Fi:diagrams} shows
 the possible types for charged and neutral $B$ decays. In \babar, the large 
data sets now available allow comprehensive investigations of these 
transitions. In the analysis described in this note, we present measurements 
of or limits on the branching fractions for all the possible 
$B\to \overline D^{(*)} D^{(*)} \KS$ and 
$B\rightarrow \overline D^{(*)} D^{(*)} K^+$ decay modes, using events in 
which both $D$ mesons are fully reconstructed. 

\begin{figure}[H]
\begin{center}
\scalebox{.29}{\includegraphics{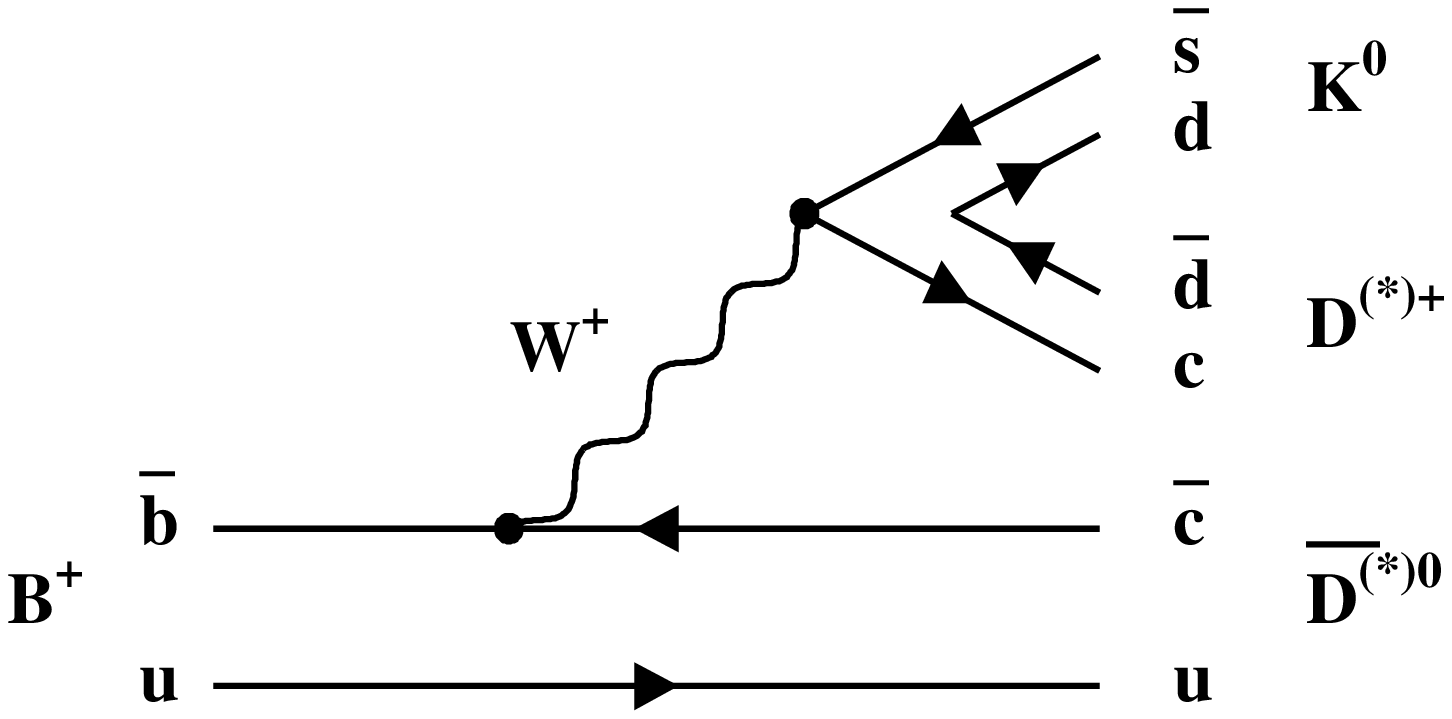}
\includegraphics{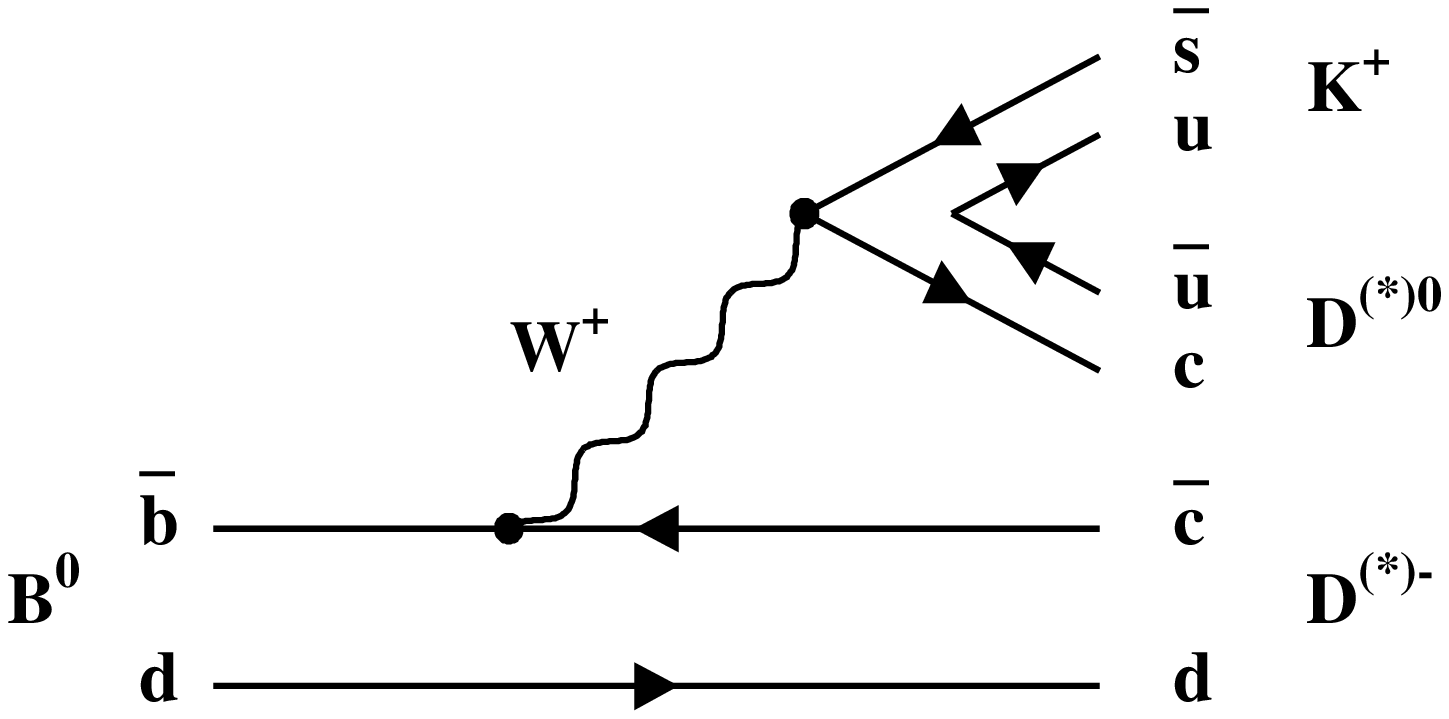}}
\end{center}
\end{figure}

\begin{figure}[H]
\begin{center}
\scalebox{.29}{\includegraphics{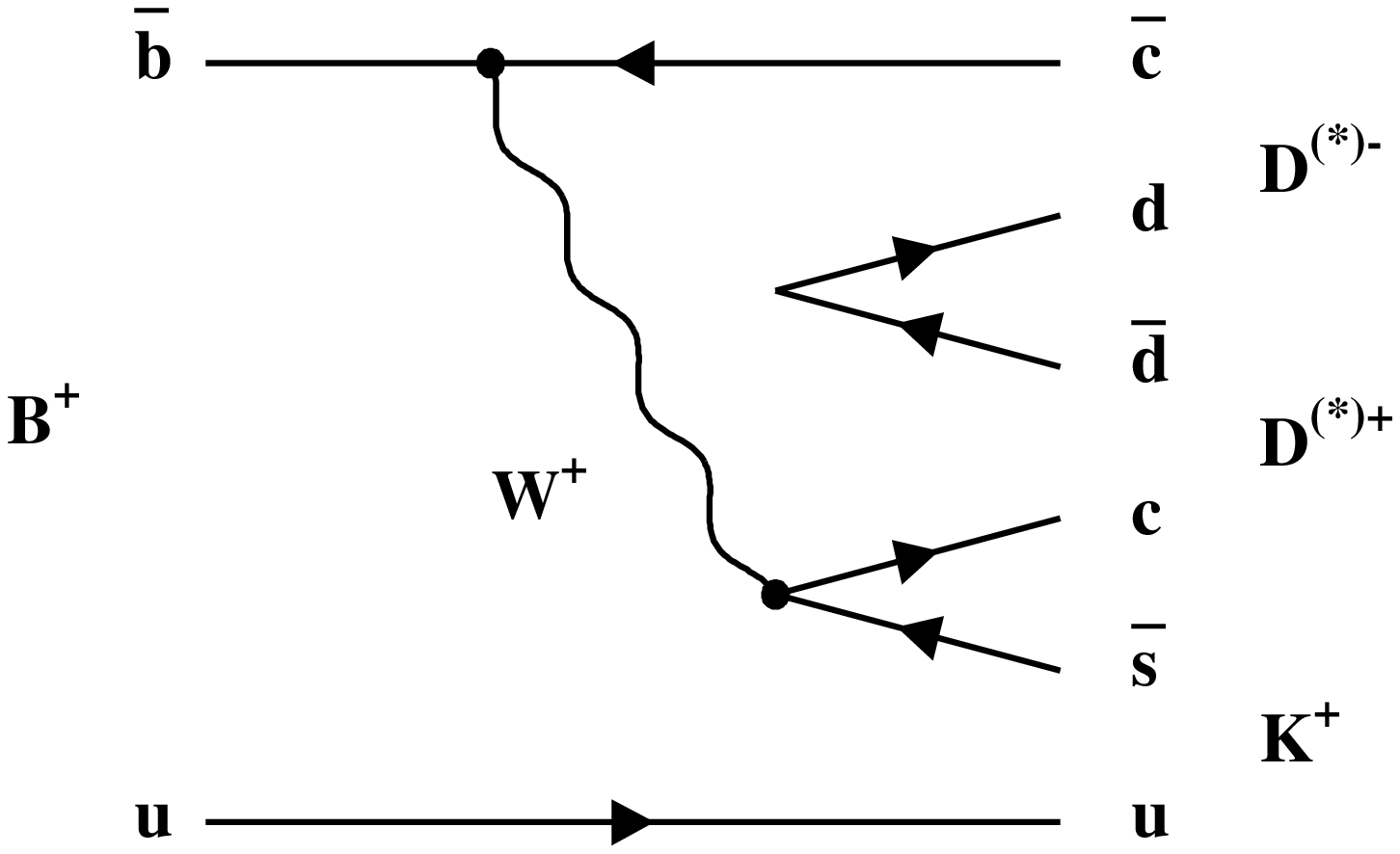}
\includegraphics{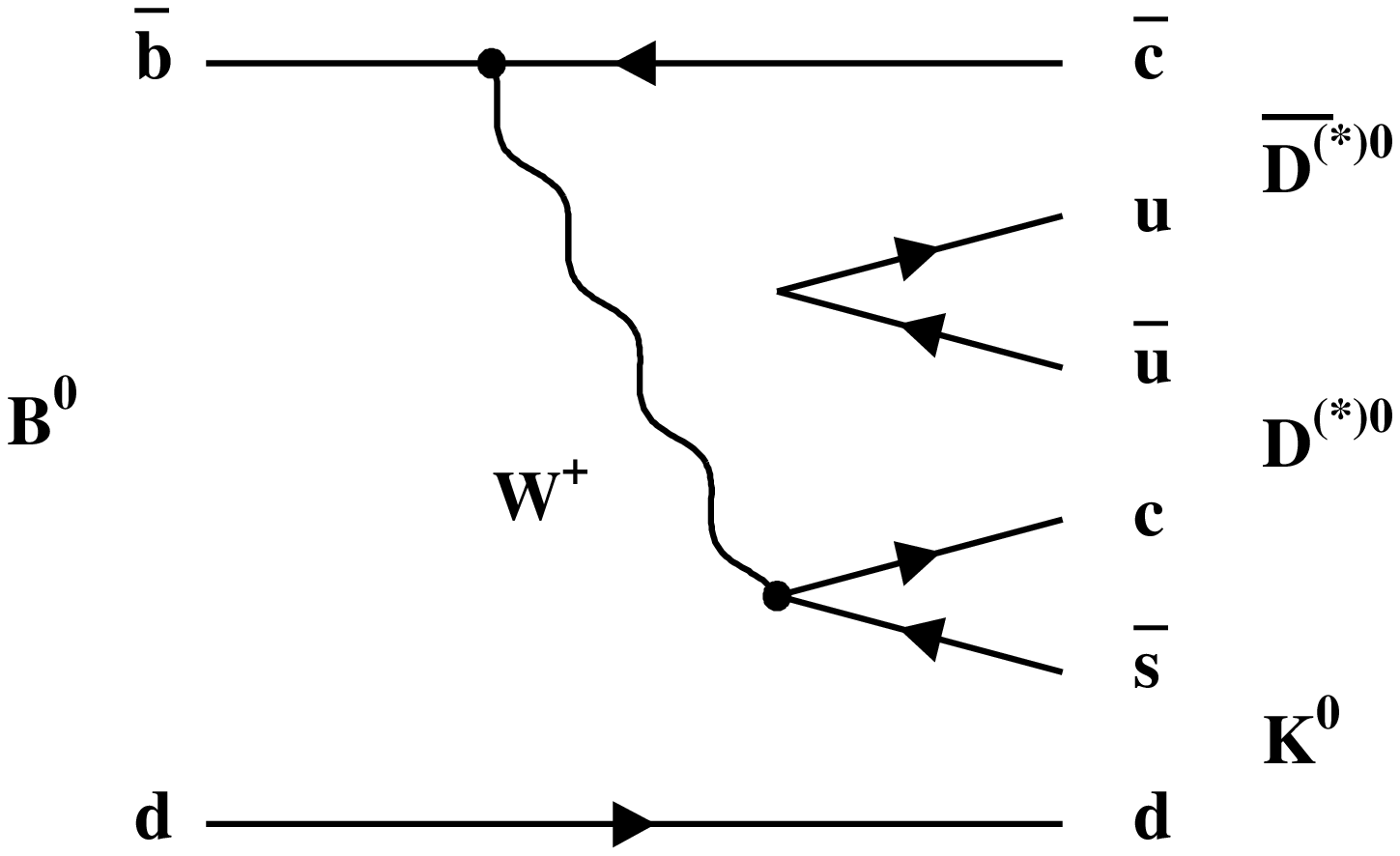}}
\end{center}
\end{figure}

\section{\boldmath The \babar\ detector and dataset}
\label{sec:babar}
The study reported here uses 75.9\invfb\ of data collected at the 
$\Upsilon(4S)$ resonance  with the \babar\ detector, corresponding to 
 $(82.3 \pm 0.9)\times 10^6$ $B \overline B$ pairs. 
\par The \babar\ detector is a large-acceptance solenoidal 
spectrometer (1.5 T) described in detail elsewhere~\cite{ref:babar}. 
The analysis described below makes use of charged track and 
$\pi^0$ reconstruction and charged particle identification.
Charged particle trajectories are measured by a 5-layer double-sided 
silicon vertex tracker (SVT) and a 40-layer drift chamber (DCH), which also 
provide ionisation measurements (\dedx) used for particle 
identification. Photons and electrons are measured in the electromagnetic 
calorimeter (EMC), made of 6580 thallium-doped CsI crystals constructed 
in a non-projective barrel and forward endcap geometry. Charged $K/\pi$ 
separation up to 4\gevc\ in momentum is provided by a detector of internally 
reflected Cherenkov light (DIRC), consisting of 12 sectors of quartz 
bars that carry the Cherenkov light to an expansion volume filled with water
and equipped with 10751 photomultiplier tubes. 

\section{\boldmath ${B}$ candidate selection}
\label{sec:Analysis}
The $B^0$ and $B^+$ mesons are reconstructed in a sample of multihadron events 
for all the possible $\overline D D K$ modes, namely 
$B^0\rightarrow D^{(*)-}D^{(*)0}K^+$, 
$D^{(*)-}D^{(*)+}K^0$, $\overline{D}^{(*)0}D^{(*)0}K^0$ and 
$B^+\rightarrow \overline{D}^{(*)0}D^{(*)+}K^0$, 
$\overline{D}^{(*)0}D^{(*)0}K^+$, $D^{(*)-}D^{(*)+}K^+$. $K^0$ mesons are reconstructed 
only from the decays $K^0_S \to \pi^+ \pi^-$. To eliminate the background from 
continuum $e^+e^- \to q \overline q$ events, we require that the ratio of 
the second to zeroth Fox-Wolfram moments \cite{ref:fox} be less that 0.45. 

The \KS candidates are reconstructed from two oppositely charged tracks
consistent with coming from a common vertex and having an invariant mass 
within $\pm 9 \mevcc$ of the nominal $K^0_S$ mass. For most of the channels 
involving a \KS, we require that the \KS vertex is displaced from the 
interaction point for the event by at least 0.2 cm in the plane 
transverse to the beam axis direction.
  The \piz candidates are reconstructed from pairs of photons, each with 
an energy greater than 30\mev, which are required to have a mass 
$115<M_{\gamma \gamma}<150\mevcc$. The \piz from \Dstarz must 
have a momentum $70<p^*(\gamma\gamma)<450\mevc$ in the $\Upsilon(4S)$ frame, 
while the \piz from $D^0\rightarrow K^-\pi^+\pi^0$ must have an energy $E(\piz)>200\mev$. 

$D^*$ candidates are reconstructed in the decay modes 
$D^{*+}\rightarrow D^0\pi^+$, $D^{*0}\rightarrow D^0\pi^0$ and    
$D^{*0}\rightarrow D^0\gamma$.
A $\pm 3 \sigma$ interval around the nominal mass difference 
$\Delta M= M(D^*)-M(D^0)$ is used to select $D^*$ mesons, where $\sigma$ is 
the measured mass resolution. For decays 
$B^0 \to D^{*-}D^{*+}K^0_S$ and $B^+ \to D^{*-}D^{*+}K^+$, one of the 
$D^{*\pm}$ is also allowed to decay to $D^\pm\pi^0$. 

The $D^0$ and $D^+$ mesons are reconstructed in the decay modes 
$D^0\rightarrow K^-\pi^+$, $K^-\pi^+\pi^0$, $K^-\pi^+\pi^-\pi^+$ and 
$D^+\rightarrow K^-\pi^+\pi^+$, by selecting track
combinations with invariant mass within $\pm 2\sigma$ of the average 
measured $D$ mass.  The average $D$ mass and the $D$ mass resolution 
$\sigma$ used in this selection
 are fitted from the data itself, using an inclusive sample of $D$ decays.
For modes involving two $D^0$ mesons, at least one of them is required to 
decay to $K^- \pi^+$, except for the decay modes $D^{*-}D^{*+}K^0$, 
$D^{*-}D^{*+}K^+$ and $D^{*-}D^0K^+$, which have lower background.  
The $K$ and $\pi$ tracks are required to be well 
reconstructed in the tracking detectors and to originate from a common vertex. 
Charged kaon identification, based on the measured Cherenkov angle in the DIRC
 and the \dedx\ measurements in the drift chamber and the vertex tracker, is 
used for most D decay modes, as well as for the $K^+$ from the $B$ meson decay.

$B$ candidates are reconstructed from one $\overline{D}^{(*)}$, one $D^{(*)}$ 
and one $K$ candidate. A mass constraint is applied to all the intermediate 
particles ($D^{*0}$, $D^{*+}$, $D^0$, $D^+$, \KS, \piz). 
Since the $B$ mesons are produced via \epem 
$\rightarrow$ \upsbb, the energy of the $B$ in the $\Upsilon{( 4S)}$ frame 
is given by the beam energy $E_{beam}^*$, which is known much more 
precisely than the energy of the $B$ candidate. Therefore, to isolate the 
$B$ meson signal, we use two kinematic variables: \de, the difference 
between the reconstructed energy of the $B$ candidate and the beam energy in 
the center of mass frame, and \mes, the beam energy substituted mass, 
defined as 
\begin{equation}
\mes = \sqrt{E_{beam}^{*2}-p_B^{*2}},
\end{equation}
where $p_B^*$ is the momentum of the reconstructed $B$ in the $\Upsilon{( 4S)}$ 
frame. Signal events have \de\ close to 0 and \mes close to the $B$ meson 
mass, 5.729\gevcc. When several candidates are selected in an event,  
only the candidate with the lowest $|\Delta E |$  value is considered 
(``best candidate''). 
From Monte Carlo studies, this algorithm is found to give the best 
reconstruction efficiency and the lower cross-feed between the different 
$\overline D^{(*)}D^{(*)}K$ modes; it is found to introduce 
no bias on the signal extraction, since the latter is performed from the \mes\
 spectra only. However, in the Fig.~\ref{fig:figsum}, to avoid the bias on 
$\Delta E$ inherent to the method, all the \de\ spectra are shown without this
 requirement.

\section{Evidence for a signal in the sum of all \boldmath ${B}$ submodes}
\label{sec:Sumresults}

\begin{figure}[!htb]
\begin{center}
\includegraphics[height=18cm]{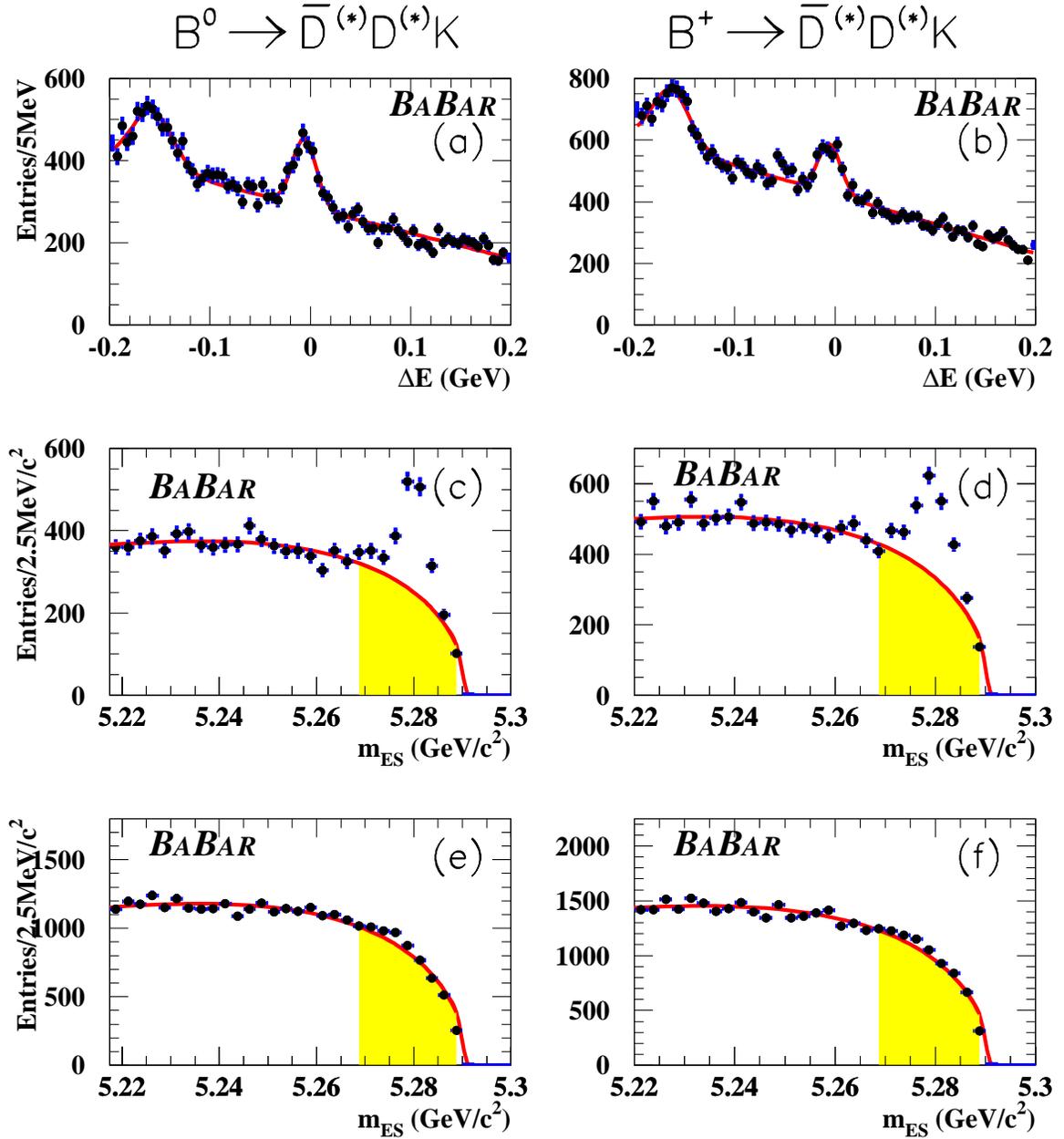}
\caption{The $\Delta E$ and $m_{\rm ES}$ spectra (a,c,e) for the sum of all the $B^0\rightarrow \overline D^{(*)}D^{(*)}K$ modes  and (b,d,f) for the sum of all the $B^+\rightarrow \overline D^{(*)}D^{(*)}K$  modes. 
 (a,b): $\Delta E$ for $m_{\rm ES}>5.27\gevcc$. 
 (c,d): $m_{\rm ES}$ for $\Delta E<2.5\sigma$ (signal box). 
 (e,f): $m_{\rm ES}$  for $\Delta E>50\mev$ (background control region). 
 The curves superimposed to the $m_{\rm ES}$ spectra correspond to the ARGUS background fits described in the text and
 the shaded regions represent the background under the signal region 
$5.27<m_{\rm ES}<5.29\gevcc$.} 
\label{fig:figsum}
\end{center}
\end{figure}

The \mes\ and  \de\ spectra of the selected events are shown in 
Fig.~\ref{fig:figsum} for the sum of all the decay modes, separately for 
$B^0$ and $B^+$. The \de\ spectra  are shown for
events in the signal region defined by $5.27<m_{\rm ES}<5.29 \gevcc$. 
Signal events appear in the peak near 0\mev\ when reconstructed correctly, 
while the peak around $-160$\mev\ is due to 
$\overline D^*DK$  or $\overline D^*D^*K$ decays reconstructed as 
$\overline D D K$  or $\overline D^*DK$, respectively.
The \mes\ spectra for the signal region are shown for events with \de\ within 
$\pm 2.5\sigma_{\Delta E}$ of the central $\Delta E$ value for the signal. The
 resolution $\sigma_{\Delta E}$ is determined from the data and is equal to 
9.9\mev\ for events involving no $D^{*0}$ and 11.3\mev\ for events involving 
one $D^{*0}$. For events with two $D^{*0}$, 
the resolution is estimated from the Monte Carlo simulation to be 13.8\mev. 
A shift $\Delta E_{\mathrm{shift}}=(-5 \pm 1)\mev$ of the \de\ central value 
for the signal is observed in the data. This shift is due to imperfect modeling
 of the charged $K$ energy losses in the detector material and is accounted 
for in the analysis. As explained above, only the candidate with the lowest 
$|\Delta E - \Delta E_{\mathrm{shift}}|$ appears in the \mes\ spectra in case 
of multiple candidates. Both the \mes\ spectra for the \de\ signal region and 
the \de\ spectra show clear evidence of a signal. 
On the contrary, the \mes\ spectra for the background control region 
$\Delta E>50\mev$ show no evidence
 of any excess of events in the $B$ signal region. For the \mes\ spectra, the 
combinatorial background is empirically described by the ARGUS function 
\cite{ref:argus}
\begin{equation}
{ dN \over dm_{\rm ES} } = f( m_{\rm ES};A,\zeta) = A \times m_{\rm ES} \times 
 \sqrt{1-{m_{\rm ES}^2 \over E_{beam}^{*2}}} \times 
 \exp{\left[-\zeta \left( 1-{m_{\rm ES}^2 \over E_{beam}^{*2}}\right)\right]},
\end{equation}
 where A is a normalisation factor. The function depends on a free parameter 
 $\zeta$ that is determined from a fit to the \mes\ spectra 
of the background control region. The number of combinatorial background 
events in the signal region is then estimated by normalizing the ARGUS 
function to the region $5.22<m_{\rm ES}<5.27\gevcc$ in the projection 
containing the signal (Fig.~\ref{fig:figsum}c,d) and extrapolating it  
to the signal region $5.27<m_{\rm ES}<5.29\gevcc$. 
The fitted ARGUS functions are overlaid on the \mes\ spectra of 
Fig.~\ref{fig:figsum}. The average number of background events 
expected in the signal region  is  1889$\pm$24 for neutral $B$  mesons and  
2512$\pm$27 for charged $B$ mesons, while 2712 and 3482 events are 
observed, giving an excess of 823$\pm$57 $B^0$ and 969$\pm$65 $B^+$ events 
in the signal region.

\section{Measurement of exclusive branching fractions}
\label{sec:exclusive}

In the following, the subscript $k$ will be used to identify  the different 
\btoddk\  decay modes (i.e., $\overline D^0 D^0K^+$, $D^{*-}D^0K^+$, ...). 
The subscript $i$ will be used to identify the different decay submodes of the 
$\overline D D$ pair (i.e., $i$ =$ K\pi\times K\pi$, $K\pi\times K\pi\pi^0$, 
$K\pi\times K3\pi$, ...). The subscript $ik$ will therefore refer to the 
$B$ mode $k$ decaying into the $\overline D D$ submode $i$. 

\begin{figure}[!htb]
\begin{center}
\includegraphics[height=18cm]{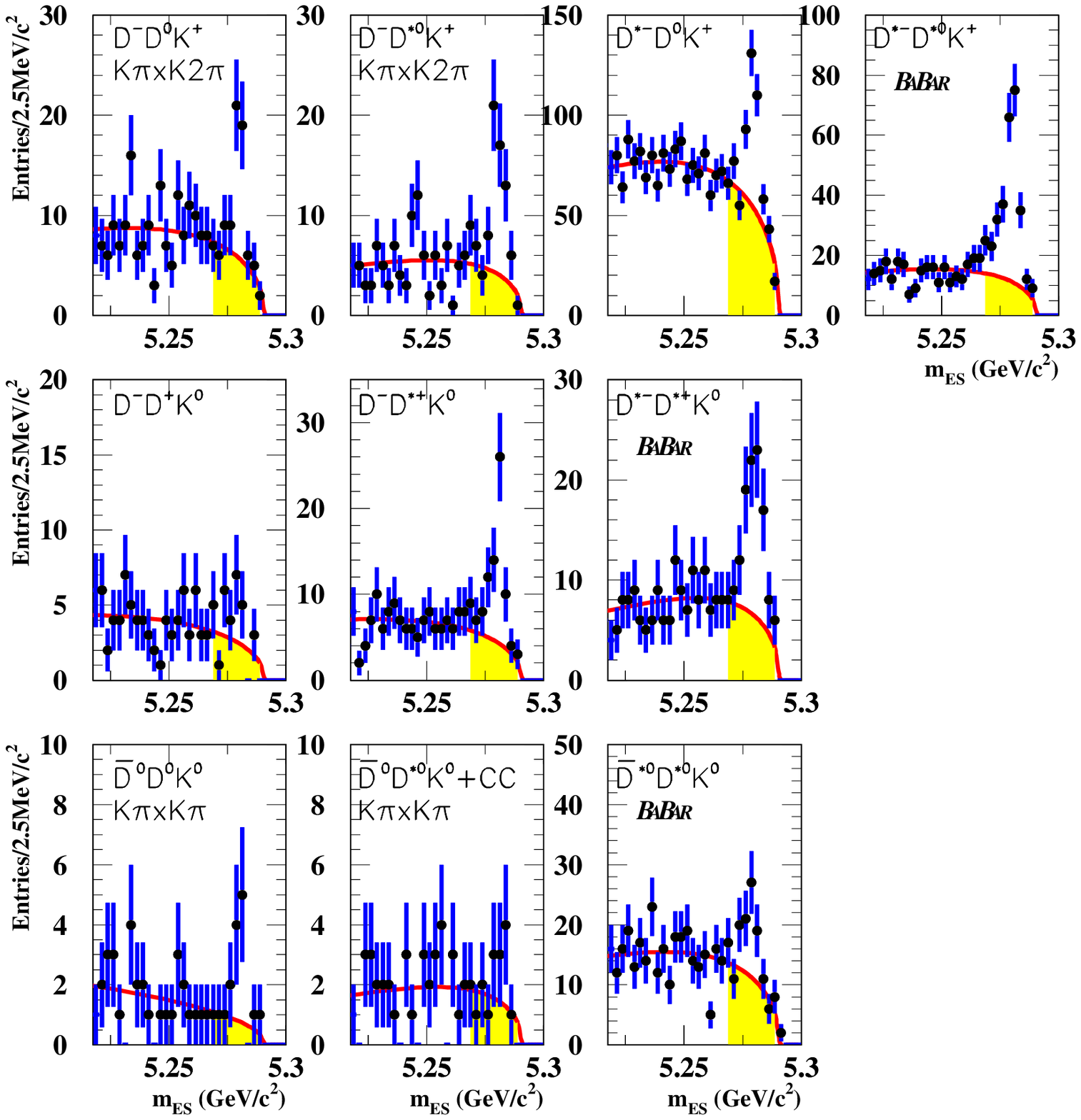}
\caption{The $m_{\rm ES}$ spectra of the ten 
$B^0\rightarrow \overline D^{(*)} D^{(*)}K$ modes. For each mode, all the $D$ 
decay submodes used in the analysis have been summed, except for plots where 
 the $\bar D \times D$ decay modes used appear explicitly. 
The curves correspond to the background fits described in the text and the 
shaded regions represent the background under the signal box.  
Upper row: pure external spectator 
$B^0\rightarrow D^{(*)-} D^{(*)0}K^+$ decays. 
Middle row: external+internal decays 
$B^0\rightarrow D^{(*)-} D^{(*)+}K^0_S$. 
Bottom row: pure internal (color suppressed) decays $B^0\rightarrow \overline D^{(*)0} D^{(*)0}K^0_S$.} 
\label{fig:b0modes}
\end{center}
\end{figure}

\begin{figure}[!htb]
\begin{center}
\includegraphics[height=18cm]{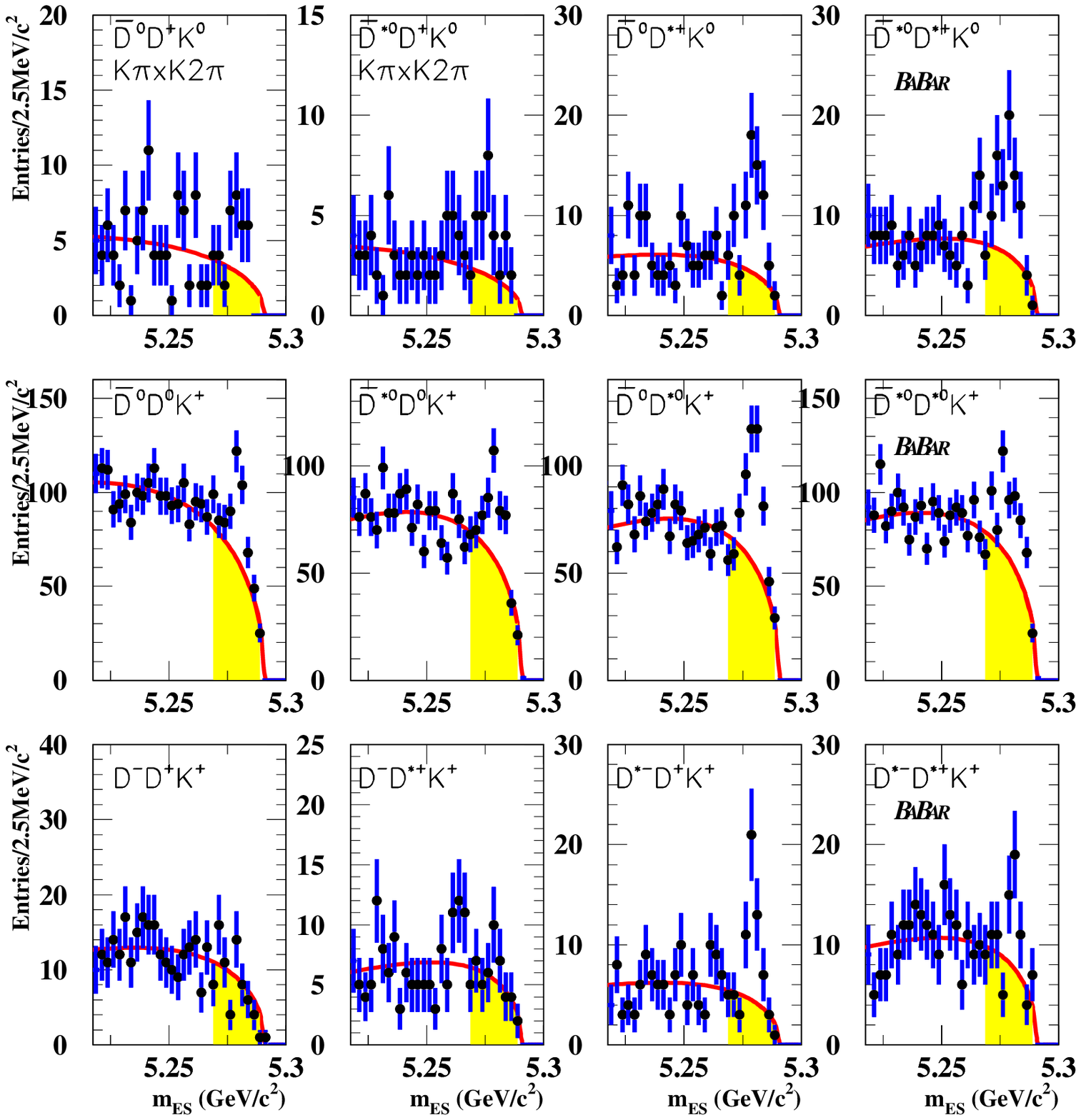}
\caption{The $m_{\rm ES}$ spectra of the twelve $B^+\rightarrow \overline D^{(*)}D^{(*)}K$ modes. 
For each mode, all the $D$ decay submodes used in the analysis have been 
summed, except for plots where the $\bar D \times D$ decay modes used appear 
explicitly. The curves correspond to the background fits described in 
the text and the shaded regions represent the background under the signal box. 
Upper row: pure external spectator decays 
$B^+\rightarrow \overline D^{(*)0} D^{(*)+}K^0_S$. Middle row: external+internal decays 
$B^+\rightarrow \overline D^{(*)0} D^{(*)0}K^+$. Bottom row: pure internal
(color suppressed) decays $B^+\rightarrow D^{(*)-}  D^{(*)+}K^+$.} 
\label{fig:bchmodes}
\end{center}
\end{figure}

The \mes\ spectra obtained after a $\pm 2.5 \sigma_{\Delta E}$ selection on 
$\left(\Delta E-\Delta E_{\mathrm{shift}}\right)$ for all the different $\overline D^{(*)} D^{(*)} K$ modes are shown in 
Fig.~\ref{fig:b0modes} ($B^0$ decay modes) and Fig.~\ref{fig:bchmodes} 
($B^+$ decay modes).
The corresponding event yields, computed as explained below, are given in 
Table \ref{tab:ddkyields}. In Figs.~\ref{fig:b0modes} and \ref{fig:bchmodes} 
and in Table \ref{tab:ddkyields}, for a given $B$ decay mode  the signals
 from the different $\overline D D$ decay submodes have been summed. 
However, to take advantage of the different signal-to-background ratio 
of the various submodes, the information from each submode is entered 
 separately in a likelihood function used to calculate the 
\btoddk\ branching fractions. 
As a first step, the ARGUS shape parameter of each submode, 
$\zeta_{ik}$, is determined from a fit to the \mes\ spectra 
of the background control region $\Delta E>50\mev$. An ARGUS function with 
the shape parameter $\zeta$ fixed to this value is then fitted
 to  the  \mes\ distribution for the signal region 
$|\Delta E-\Delta E_{\mathrm{shift}}|<2.5\sigma_{\Delta E}$, 
excluding from the fit events with $5.27 < m_{\rm ES} < 5.29\gevcc$. 
A value for the background normalization parameter $A_{ik}$ is calculated and 
the number of background events $N^{\rm bkg}_{ik}$ in the signal region for this 
submode is calculated as  
\begin{equation}\label{Eq:nbkg}
N^{\rm bkg}_{ik} = \int_{5.27}^{5.29} f\left(x;A_{ik},\zeta_{ik}\right) dx.
\end{equation}
If $n_k$ submodes are used for a given mode, the branching fraction for that mode is then extracted by maximizing the following likelihood: 
\begin{equation}\label{Eq:poisson}
L_k = \prod_{i=1}^{n_k} \frac{{\mu_{ik}}^{N_{ik}} e^{-\mu_{ik}}}{N_{ik}!},
\end{equation}
where $N_{ik}$ is the observed number of events in the signal 
region and $\mu_{ik}$ is the predicted number of events in the signal region. 
$\mu_{ik}$ is the sum of three contributions:
\begin{itemize}
\item the predicted signal $N_{ik}^S$, which is related to
the (unknown) branching fraction ${\cal B}_k$ of decay mode $k$, 
the reconstruction efficiency ($\epsilon_{ik}$), the intermediate branching fractions 
${\cal B}_{i}^{\overline D D}$ and the number of $B\overline{B}$ events ($N_{B\overline{B}}$)
\begin{equation} 
N_{ik}^S = {\cal B}_k \times N_{B\overline{B}} \times \epsilon_{ik} \times {\cal B}_{i}^{\overline D D};
\end{equation}

\item the number of combinatorial background events $N^{\rm bkg}_{ik}$, determined as described 
above (Eq.\ref{Eq:nbkg});
\item the peaking background $N^{\rm peak}_{ik}$ from other \btoddk\ decay modes, calculated as 

\begin{equation}
N_{ik}^{\rm peak} = \sum_{l \ne k}  {\cal B}_l \times N_{B\overline{B}} \times \epsilon '(il\to ik)  \times {\cal B}_{i}^{\overline D D},
\end{equation}
where $\epsilon '(il\to ik)$ is the cross-feed matrix from $B$ mode $l$ to $B$ mode 
$k$ for the $\overline D D$ decay submodes $i$ (the cross-feed between different 
$\overline D D$ decay submodes is found to be negligible). The only significant 
cross-feed is observed between decay modes where a fake $D^{*0}$ replaces a true 
$D^{*+}$ or a true $D^0$, for instance between $D^{*-} D^0 K^+$ and 
$\overline D^{*0} D^{0} K^+$, or between $\overline D^{*0} D^{0}K^+$ and  $\overline D^0 D^{*0}K^+$. 
Therefore, these branching fractions are extracted with joint likelihood 
 in Eq.~\ref{Eq:poisson}.
\end{itemize} 
The $D^*$ and $D$ branching fractions used in the branching fraction 
calculation are summarized in Table \ref{tab:brpdg} \cite{ref:pdg}.
The selection efficiencies and the cross-feed matrices for each mode are 
obtained from a detailed Monte Carlo simulation, in which the detector 
response is modeled with the GEANT4 program. $B$ meson decays to 
$\overline DDK$ are generated with a three-body phase space model in the 
simulated event samples used for the efficiency calculation. 
For each decay submode, samples of about  15000 signal events have been 
produced. In addition, data are used whenever 
possible to determine detector performance: tracking efficiencies are 
determined by identifying tracks in the silicon vertex detector and measuring 
the fraction that is well reconstructed in the drift chamber; the kaon 
identification efficiency is estimated from a sample of 
$D^{*+}\rightarrow D^0\pi^+$, $D^0\rightarrow K^-\pi^+$ decays; the $\gamma$ 
and $\pi^0$ efficiencies are measured by comparing the ratio of events 
$N(\tau^+ \to \overline \nu_{\tau} h^+\pi^0)/N(\tau^+ \to \overline \nu_{\tau} h^+\pi^0\pi^0)$ to the previously 
measured branching fractions \cite{cleotau}.
 Typical efficiencies range from 20\%, 
for \bdzdzk\ with both $D^0$ mesons decaying to $K^-\pi^+$,  to less than 1\%, 
for \btodsdsk\ ($D^{*+} \to D^0 \pi^+$, $D^{*-} \to \overline D^0 \pi^-$ ) 
with $D^0$ mesons decaying to $K^-\pi^+\pi^0$ or $K^-\pi^+\pi^-\pi^+$.
\begin{table}[!htb]
\caption{Number of events and branching fractions for each mode. The first 
error on the branching fraction is the statistical uncertainty and the second 
one is the systematic uncertainty.}
\begin{center}
\begin{tabular}{|l|c|c|c|c|c|} \hline
            & Total &            &         & Branching & 90\%~C.L.\\
  B decay   & yield in & Estimated  &         & fraction  & upper   \\
    mode    & signal region & background & Excess  & $\left(10^{-3}\right)$& limit $\left(10^{-3}\right)$  \\ \hline
\multicolumn{6}{|c|}{$B^0$ decays through external W-emission amplitudes} \\ \hline
 $B^0\rightarrow D^{-} D^{0}K^+$ & 599 & 479$\pm$12 & 120$\pm$27 & $1.7 \pm 0.3 \pm 0.3$ &\\
$B^0\rightarrow D^{-} D^{*0}K^+$ & 468 & 337$\pm$10 & 131$\pm$24 & $4.6 \pm 0.7 \pm 0.7$ &\\
$B^0\rightarrow D^{*-} D^{0}K^+$ & 584 & 399$\pm$11 & 185$\pm$27 & $3.1^{+0.4}_{-0.3}\pm 0.4$ &\\
$B^0\rightarrow D^{*-} D^{*0}K^+$ & 289 & 84$\pm$5  & 205$\pm$18 & $11.8 \pm 1.0 \pm 1.7 $&\\ \hline
\multicolumn{6}{|c|}{$B^0$ decays through external+internal W-emission amplitudes} \\ \hline
$B^0\rightarrow D^{-} D^{+}K^0$ & 26 & 19$\pm$2 & 7$\pm$5 & $0.8^{+0.6}_{-0.5}\pm 0.3$ & $<1.7$\\ 
$B^0\rightarrow D^{*-} D^{+}K^0$+CC & 84 & 34$\pm$3 &50$\pm$10 & $6.5 \pm 1.2 \pm 1.0$ & \\ 
$B^0\rightarrow D^{*-} D^{*+}K^0$ &116 & 48$\pm$4 &68$\pm$11 & $8.8^{+1.5}_{-1.4}\pm 1.3$ &\\ \hline
\multicolumn{6}{|c|}{$B^0$ decays through internal W-emission amplitudes} \\ \hline
$B^0\rightarrow \overline D^{0} D^{0} K^0$ & 175 & 173$\pm$7 & 2$\pm$15 & $0.8 \pm 0.4 \pm 0.2$ & $<1.4$ \\ 
$B^0\rightarrow \overline D^{0} D^{*0} K^0$+CC & 248 & 225$\pm$8 & 23$\pm$18 & $1.7^{+1.4}_{-1.3}\pm 0.7$ & $<3.7$ \\ 
$B^0\rightarrow \overline D^{*0} D^{*0} K^0$& 123 &  81$\pm$6 & 42$\pm$13 & $3.3^{+2.1}_{-2.0}\pm 1.4$ & $<6.6$ \\ \hline
\multicolumn{6}{|c|}{$B^+$ decays through external W-emission amplitudes} \\ \hline
$B^+\rightarrow \overline D^{0} D^{+}K^0$ & 367 & 317$\pm$9 & 50$\pm$21 & $1.8 \pm 0.7 \pm 0.4$ & $<2.8$ \\
$B^+\rightarrow \overline D^{*0} D^{+}K^0$ & 216 & 175$\pm$7 & 41$\pm$16 & $4.1^{+1.5}_{-1.4}\pm 0.8$ & $<6.1$ \\
$B^+\rightarrow \overline D^{0} D^{*+}K^0$ &  77 & 31$\pm$3 & 46$\pm$9 & $5.2^{+1.0}_{-0.9}\pm 0.7$ &\\
$B^+\rightarrow \overline D^{*0} D^{*+}K^0$ & 89 & 43$\pm$4 & 46$\pm$10 &$7.8^{+2.3}_{-2.1}\pm 1.4$ & \\ \hline
\multicolumn{6}{|c|}{$B^+$ decays through external+internal W-emission amplitudes} \\ \hline
$B^+\rightarrow \overline D^0 D^0 K^+$     & 627 & 469$\pm$11 &158$\pm$27 & $1.9 \pm 0.3 \pm 0.3$ &\\
$B^+\rightarrow \overline D^{*0} D^{0}K^+$ & 552 & 411$\pm$11 &141$\pm$26 & $1.8^{+0.7}_{-0.6} \pm 0.4$ & $<3.8$ \\ 
$B^+\rightarrow \overline D^{0} D^{*0}K^+$ & 623 & 402$\pm$11 &221$\pm$27 & $4.7 \pm 0.7 \pm 0.7$ &\\ 
$B^+\rightarrow \overline D^{*0} D^{*0}K^+$ & 675 & 468$\pm$15 & 207$\pm$30 & $5.3^{+1.1}_{-1.0} \pm 1.2$ &\\  \hline
\multicolumn{6}{|c|}{$B^+$ decays through internal W-emission amplitudes} \\ \hline
$B^+\rightarrow D^{-} D^{+}K^+$ & 64 & 65$\pm$4 & -1$\pm$9 & $0.0 \pm 0.3 \pm 0.1$ & $<0.4$ \\
$B^+\rightarrow D^{-} D^{*+}K^+$ & 45 & 39$\pm$4 & 6$\pm$8 & $0.2 \pm 0.2 \pm 0.1$ & $<0.7$ \\
$B^+\rightarrow D^{*-} D^{+}K^+$ & 64 & 32$\pm$3 & 32$\pm$9& $1.5 \pm 0.3 \pm 0.2$ &  \\
$B^+\rightarrow D^{*-} D^{*+}K^+$ & 83 & 60$\pm$4 & 23$\pm$10 & $0.9 \pm 0.4 \pm 0.2$ & $<1.8$ \\ \hline
\end{tabular}
\end{center}
\label{tab:ddkyields}
\end{table}

\begin{table}[!htb]
\caption{Submode branching fractions used in the analysis \cite{ref:pdg}. 
The errors on the ${\cal B}(D^0\to K^- \pi^+ \pi^0)$ and ${\cal B}(D^0 \to K^- \pi^+ \pi^- \pi^+)$ 
correlated to the error on ${\cal B}(D^0 \to K^- \pi^+)$ are indicated separately with the subscript $K\pi$.}
\begin{center}
\begin{tabular}{|l|c|}
\hline
\textbf{Mode} & ${\cal B}$ (\%)  \\ 
\hline\hline
$D^0\to K^- \pi^+$ & $3.80\pm 0.09\ $ \\ 
$D^0\to K^- \pi^+ \pi^0$ &$13.10 \pm 0.84 \pm 0.31_{K\pi}$ \\ 
$D^0 \to K^- \pi^+ \pi^- \pi^+$ & $7.46  \pm 0.30 \pm 0.18_{K\pi}$\\ \hline
$D^+\to K^- \pi^+ \pi^+$ & $9.1 \pm 0.6\ $ \\ \hline
$D^{*+}\to D^0 \pi^+$ & $67.7 \pm 0.5\ $ \\
$D^{*+}\to D^+ \pi^0$ & $30.7 \pm 0.5\ $ \\ \hline
$D^{*0}\to D^0 \pi^0$ & $61.9 \pm 2.9\ $ \\ 
$D^{*0}\to D^0 \gamma$ &$38.1 \pm 2.9\ $ \\ \hline
$K^0_S \to \pi^+ \pi^-$&$68.61 \pm 0.28\ $ \\ \hline
\end{tabular}
\end{center}
\label{tab:brpdg}
\end{table}

\section{Systematic studies}
\label{sec:Systematics}
Due to the large number of $K^\pm$ and to the large multiplicities involved in
the decays \btoddk, the dominant systematic uncertainties come from our level 
of understanding of the charged kaon identification and of the charged 
particle tracking efficiencies. Both systematic uncertainties are estimated on a per track
 basis and are given in Table \ref{tab:systematics}. Other systematic 
uncertainties are due to uncertainties on the $D$ and $D^*$ branching 
fractions, the $\pi^0$ reconstruction efficiencies, the $D$ vertexing 
requirements, and the $\Delta E$ resolution used to define the 
signal box, as well as the uncertainty on the combinatorial background 
estimates, the statistical uncertainty on the efficiency due to the finite 
size of the Monte Carlo simulation samples and the uncertainty on 
the number of $B \overline B$ events in the data sample. The fractional 
systematic uncertainties on efficiencies and the branching fractions 
are summarized in Table \ref{tab:systematics}.
\par Possible decay model dependence of the efficiencies were also studied by 
generating decays $B^0 \rightarrow D^{*-}D_{s1}^+$ and 
$B^0 \rightarrow D^{*-}D'^+_{s1}$
($D^+_{s1}, D'^+_{s1}\rightarrow D^{*0}K^+)$, 
where $D^+_{s1}$ is the narrow ($\Gamma=1\mev$, $M=2.536\gevcc$) orbitally 
excited  $1^+$ state of the $D^{**}_s$ system and $D'^+_{s1}$ is a wide ($\Gamma=250\mev$, $M=2.560\gevcc$)  
$D^{**}_s$ resonance. The efficiency for reconstructing these 
modes was compared to the efficiency found for decays 
$B^0 \rightarrow D^{*-} D^{*0}K^+$ generated with a phase space model. 
 We found no statistically significant difference in efficiencies; we assign 
a systematic uncertainty equal to the statistical error of 5\%. 

\begin{table}[!htb]\caption{Fractional systematic uncertainties 
on efficiencies and branching fractions.}
\begin{center}
\begin{tabular}{|l|l|}
\hline
Item & Fractional uncertainty on efficiency or branching fraction  \\
\hline\hline
Charged tracks reconstruction & 0.8\% per track for good tracks in Drift Chamber \\
                       & 1.2\% per track for tracks without Drift Chamber requirements \\ \hline
$K^0_S$ reconstruction        & 2.5\% per $K^0_S$, added in quadrature to the track reconstruction error \\ \hline 
$\pi^0$ reconstruction & 5.1\% per $\pi^0$   \\ \hline 
$\gamma$ from $D^{*0} \rightarrow D^0 \gamma$ & 5.1\% per $\gamma$ (correlated with the $\pi^0$ systematic) \\ \hline  
$K^\pm$ identification & 2.5\% per $K^\pm$  \\ \hline
Vertex         & 1.3\%  per 2 track vertex  \\
reconstruction & 3.1\%  per 3 track vertex  \\
               & 5.7\%  per 4 track vertex  \\ \hline 
$\sigma(\Delta E)$ & 2\% for modes with 0 or 1 $D^{*0}$ \\
                   & 5\% for modes with two $D^{*0}$'s \\ \hline  
Monte Carlo statistics & 2\% to 10\% per $\overline D D$ submode (mode and submode dependent) \\ \hline  
Intermediate br. fraction        & see Table \ref{tab:brpdg}  \\ \hline 
Number of $B \overline B$   & 1.1\%  \\ \hline 
Decay model & 5\%  \\ \hline  
\end{tabular}
\end{center}
\label{tab:systematics}
\end{table}

\section{Conclusions}
\label{sec:conclusion}

A preliminary measurement of the branching fractions for the 22 \btoddk\ modes
 is given in Table \ref{tab:ddkyields}. For the channels for which 
$\mathrm{S/\sqrt{B}}$ is smaller than 4,  a 90\% confidence level upper limit 
is also derived. (Here, B is the 
sum of the combinatorial background and of the cross-feed background from 
other $\overline D^{(*)}D^{(*)}K$ modes 
and $\mathrm{S=N-B}$, where N is the total yield in the signal region).  
This is the first time that a complete measurement of all the 
possible \btoddk\ channels is performed. The measured branching fractions 
are in good agreement with earlier measurements made with smaller data sets 
for some of these modes 
\cite{cleoddk,alephddk,babarddk,belleddk}. For the decays proceeding through 
 external W-emission or through the sum of external and internal W-emission 
 amplitudes, the branching fractions  
${\cal B}(B\rightarrow \overline D^* D K)$ and 
${\cal B}(B\rightarrow \overline D D^* K)$ are found to be about twice the 
branching fraction  ${\cal B}(B\rightarrow \overline D D K)$. The branching 
fraction ${\cal B}(B\rightarrow \overline D^* D^* K)$ is found to be about 5 
times larger than the branching fraction  
${\cal B}(B\rightarrow \overline D D K)$. No significant difference is 
observed between decays proceeding through external spectator amplitudes and 
decays proceeding through the sum of external and internal spectator 
amplitudes. 

After summing over all submodes, the preliminary branching fractions of the 
$B^0$ and of the $B^+$ to $\overline D^{(*)} D^{(*)} K$ are found to be   
\begin{equation}
{\cal B}(B^0\rightarrow \overline D^{(*)} D^{(*)} K) = \left (4.3 \pm 0.3 (stat) \pm 0.6 (syst) \right )\times 10^{-2},
\end{equation}
\begin{equation}
{\cal B}(B^+\rightarrow \overline D^{(*)} D^{(*)} K) = \left (3.5 \pm 0.3(stat) \pm 0.5 (syst) \right )\times 10^{-2}.
\end{equation}
This study confirms that a significant fraction of the transitions  
$b \rightarrow c \overline c s$ proceeds through the decays  
$B\rightarrow \overline D^{(*)} D^{(*)}K$. These decay modes account for 
 about one half of the wrong-sign $D$ production rate in $B$ decays, 
${\cal B}(B \to D\, X)=(7.9\pm 2.2)\% $ \cite{cleoupv}; however, because of 
the large statistical error on the latter measurement, it is not yet clear 
whether they saturate it. 
\par Future developments should include a search for 
$D^{**}_s$ resonant substructures in \btoddk\ decays, as well as a new high 
statistics measurement of the wrong sign $D$ production in $B$ decays and a 
search for decays $B\rightarrow \overline D^{(*)} D^{(*)}K^*$ or  
$B\rightarrow \overline D^{(*)} D^{(*)}K(n\pi)$.

\section{Acknowledgments}
\label{sec:Acknowledgments}

We are grateful for the 
extraordinary contributions of our \pep2\ colleagues in
achieving the excellent luminosity and machine conditions
that have made this work possible.
The success of this project also relies critically on the 
expertise and dedication of the computing organizations that 
support \babar.
The collaborating institutions wish to thank 
SLAC for its support and the kind hospitality extended to them. 
This work is supported by the
US Department of Energy
and National Science Foundation, the
Natural Sciences and Engineering Research Council (Canada),
Institute of High Energy Physics (China), the
Commissariat \`a l'Energie Atomique and
Institut National de Physique Nucl\'eaire et de Physique des Particules
(France), the
Bundesministerium f\"ur Bildung und Forschung and
Deutsche Forschungsgemeinschaft
(Germany), the
Istituto Nazionale di Fisica Nucleare (Italy),
the Research Council of Norway, the
Ministry of Science and Technology of the Russian Federation, and the
Particle Physics and Astronomy Research Council (United Kingdom). 
Individuals have received support from 
the A. P. Sloan Foundation, 
the Research Corporation,
and the Alexander von Humboldt Foundation.

\end{document}